%% file: lykkenlp99.tex
\documentstyle[12pt,epsfig]{article}

\catcode`\@=11 
\def\lsim{\mathrel{\mathpalette\@versim<}}
\def\gsim{\mathrel{\mathpalette\@versim>}}
\def\@versim#1#2{\vcenter{\offinterlineskip
        \ialign{$\m@th#1\hfil##\hfil$\crcr#2\crcr\sim\crcr } }}
\catcode`\@=12 
\input LP99macros.tex

\def\Title#1{\begin{center} {\Large {\bf #1} } \end{center}}

\begin{document}

\Title{Physics Needs for Future Accelerators}

\bigskip\bigskip


\begin{raggedright}  

{\it J. D. Lykken\index{Lykken, J. D.}\\
Theoretical Physics Dept.\\
Fermi National Accelerator Laboratory\\
Batavia, IL 60510}
\bigskip\bigskip
\end{raggedright}


\section{Prologomena to any meta future physics}

The title assigned to this talk, while certainly both
eye-catching and pithy, doesn't make any literal sense in English.
``Physics needs for future accelerators'' is a phrase which
could be interpreted in any number of ways, leading to very different
sorts of talks. Let me begin, therefore, by briefly describing
the roads {\it not} taken in my review. This will allow me to
mention some important issues which, while not the focus of
my talk, are worthy of serious high-minded discussion in
international forums such as the Lepton-Photon meetings.

\subsection{Physics needs for {\it building} future accelerators}

This, it seems to me, is the
most obvious literal interpretation of the original title.
Since, like all theorists, I consider myself to be
essentially omniscient, I even briefly considered giving
such a talk. I went as far as dusting off my copy
of Jackson~\cite{Jackson75}, \index{Jackson, J.D.} 
which contains, after all,
most of the basic physics that you will need to build future accelerators. 

\begin{figure}[htb]
\begin{center}
\epsfig{file=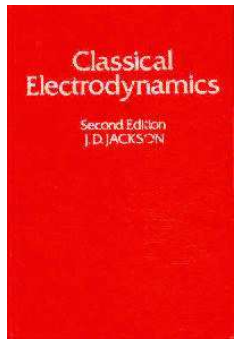,width=3cm}
\caption{The book.}
\label{fig:Redbook}
\end{center}
\end{figure}

This facetious exercise illustrates a worrisome development
to which we had better start paying more attention.
Since (in principle) all of us
understand the basics of accelerator physics, there is a
disturbing tendency to denigrate this essential branch of our
field, to relegate it to ``technicians'' --the implication
being that first-rate minds are drawn to more ``cutting-edge''
physics problems. Combined with the general trend towards
increasing specialization, we have largely decoupled and
discounted an activity which in fact largely defines and
limits the future of our field.

Furthermore, if you look seriously at
the designs and R\&D work related to any of the
proposed future accelerators
--linear colliders~\cite{TESLA98,NLC96}, \index{TESLA}\index{NLC}
hadron colliders~\cite{VLHC97}, \index{VLHC}
as well as very ambitious ideas like the
muon collider~\cite{MUCOL99}, \index{muon collider}
or the CLIC two-beam acceleration concept~\cite{CLIC99}-- \index{CLIC}
you will find
a host of interesting and highly challenging physics problems.
As a community, I doubt that we are doing enough towards
attracting, training, supporting, and encouraging
the next generation of accelerator physicists --the cadre of
first-rate, creative, and experienced people without whom
no future accelerators are likely to get built. In that case
everything else I say in this talk addresses a moot topic,
since we will have failed before we have even properly begun.

How can we avoid this calamity?
One way to attract new talent to accelerator R\&D,
and to validate the importance of this activity,
is for more experimentalists --and even theorists--
to devote some finite fraction of their time to thinking about
accelerator physics problems. 
We should keep in mind that lack of technical experience
has the positive virtue of stimulating new kinds of
questions, new ways of thinking, and thus ultimately
to innovation. 
Bringing fresh minds to bear
on these sorts of problems can only help, and will enhance
the prestige, and thus the overall vigor, of these activities.
Small steps in this direction can be encouraged now with
existing resources; a few years down the road we should aim
to establish new centers for advanced accelerator physics.

The payoff for nuturing this branch of physics will be
well worth the investment, and could be spectacular.
It may well be, for example, that physics innovation,
rather than engineering or manufacturing breakthroughs,
is the key to major cost reductions in linear collider
or hadron collider designs.

\subsection{Physics needs for {\it funding} future accelerators}

Another possible interpretation of the title of this talk is
a little more political. What do we need to get out of existing
experiments and facilities in order to strengthen the case for
particular future machines, as well as goose up the overall
momentum and glamour of our field?

\begin{figure}
\begin{center}
\epsfig{file=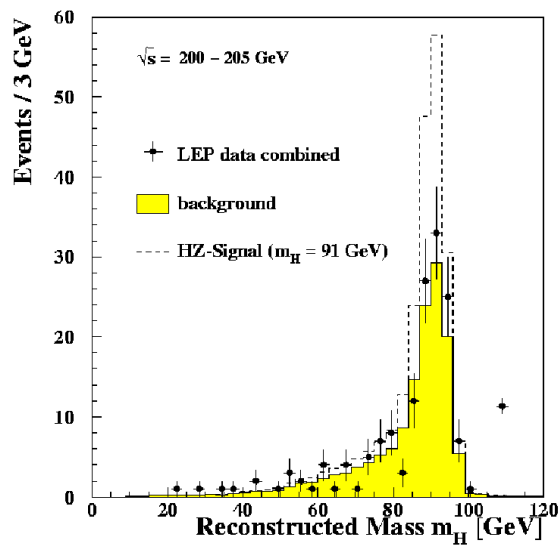,width=8cm}
\caption{A plot from LP01?}
\label{fig:FakeH}
\end{center}
\end{figure}

While we are correct to spend sleepness nights agonizing over
the optimal path to our long term future, we should also
rejoice that our immediate future looks extremely bright.
As this transparency from {\it the next} Lepton-Photon conference
illustrates, the LEP experiments have a very real chance of
discovering the Higgs. The B factories coming on line now
will provide fundamental new inputs to our global view of
particle physics, including, most likely, some surprises.
The same can be said for the many neutrino experiments in
progress or under construction, and for a number of other
low energy projects. As a bonus, a flood of new astrophysical
data will impact on a number of important ideas and
problems circulating in particle physics.

Last but not least,
the next run of the Tevatron will extend our reach for
the Higgs~\cite{HWG99}, \index{Higgs}
supersymmetry~\cite{SHW99}, \index{supersymmetry}
B physics, top physics, electroweak physics,
and new strong dynamics, to name but a few. Major discoveries
are very possible. If we are fortunate, we might even get the
first experimental hints of
extra spatial dimensions~\cite{Dimopoulos98,Lykken99},
\index{Arkani-Hamed, N.}\index{Dimopoulos, S.} \index{Dvali, G.}
\index{extra dimensions}
quantum gravity, or
strings~\cite{Lykken96,Accomando99,Cullen00}.
\index{strings, TeV}

\begin{figure}
\begin{center}
\epsfig{file=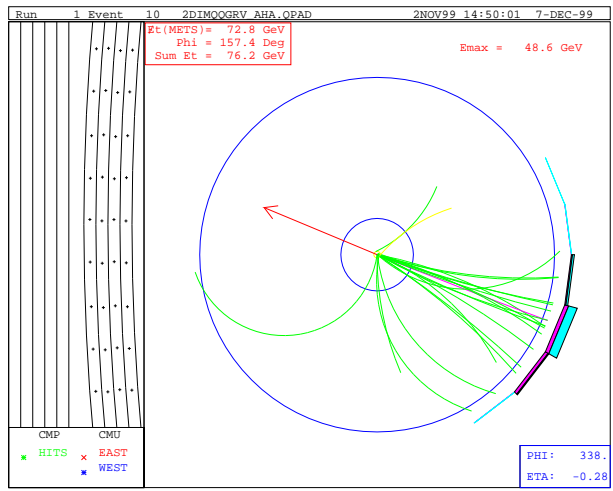,width=8cm}
\caption{Extra dimensions at the Tevatron: a monojet recoiling
off a Kaluza-Klein graviton. Simulated event from
M. Spiropulu, CDF.}
\label{fig:Monojet}
\end{center}
\end{figure}

Obviously new discoveries from any of these arenas will
help bootstrap funding and resources for future experiments
and new facilities. In addition, discoveries --or even hints--
of physics beyond the Standard Model will crystalize
our thinking about what future facilities are needed.

We can congratulate ourselves on having assembled such a rich
physics program for our immediate future. There is a danger,
however, that in our rush to get to the LHC and other future
accelerators, we may not maximally exploit those opportunities
already at hand. If we are so foolish as to sacrifice our
present to ``secure'' our future, we are more likely to end
up sacrificing {\it both}.    

\section{Physics {\it questions} for future accelerators}

What is the right way to think about future physics? The
traditional approach is to enumerate a number of possible
scenarios for physics beyond the Standard Model, guided by
whatever theoretical prejudices seem most fashionable at the moment.
Although I will myself be guilty of this sort of theorizing
later in the talk, I don't think this is the best approach
for making major decisions, decisions that will commit billions of dollars
of the taxpayers' money and determine the future or nonfuture
of high energy physics. 

I will attempt instead to outline a more robust approach to
thinking about future physics, one which relies less on
theoretical assumptions and more on an appreciation for what
high energy physics is really all about.

\subsection{Crimes and misapprehensions}

The first thing we need to do is to discredit some
harmful dogmas which have been clouding our thinking about
high energy physics for more than a decade.
Anyone who falls under the spell of these dogmas is
not likely to hold rational or constructive views
about the future needs of our field.
As far as I know, no one is willing to admit authorship
of these pernicious doctrines, but certainly the
bourgeois overlords of high energy physics and their
running dog lackeys (i.e. this audience) must assume
responsibility for tolerating or promulgating these
ideas.

The basic crime here is to have allowed the misapprehension
--among ourselves, students, and the public-- that particle
physics is ``almost done''. This misapprehension arises from
two rather different but equally radical notions, which I
will now briefly review.

\subsubsection{Organized religion}

This is the dogma that high energy physics has been
to a large extent supplanted by a new activity called
string theory. \index{string theory}
String theory is the one true
Theory of Everything (TOE), people who are a lot
smarter than you will have it figured out any day now,
and they will soon be able to compute the electron mass, etc.
purely on the basis of mathematical consistency.
Thus the traditional activities of high energy physics
(such as experiment) have become largely irrelevant.
Put another way, since no future accelerator can ever
directly probe the most fundamental scale of physics,
the ``bottom up'' approach is pointless, and we should
instead invest in the more promising ``top down''
approach to connecting fundamental physics with our
existing data banks.
   
This extreme view is, of course, quite silly. String theory
does indeed hold great promise for advancing our understanding
of fundamental physics, and has already produced some profound
insights about black hole physics, about gauge theories, and
other areas as well. But anyone who has followed the rapid
advances in string theory knows that, for every question
successfully disposed of, three new ones seem to crop up in its
wake. Fundamental physics is, not surprisingly, rich, dense,
and confusing. The road to fundamental understanding will be
a long road, and this makes the traditional activities of
high energy physics (such as experiment) even more interesting
and important in that light.

\subsubsection{Feudalism}

This is the dogma that the Standard Model is king and will
reign forever. This is particularly discouraging for young
people, because the implication here is that all the good
stuff happened in the seventies and {\it you missed it}.
There are a few things left to do --we'll find the Higgs,
measure a few more parameters-- and then that's it. So
unfortunately young physicists entering the field today
are coming in at the tail end of the Golden Age, but
--tough luck-- there's only one Golden Age and ours is almost
over.

This extreme view is equally silly, yet it seems to have
penetrated into the morose subconscious of a large fraction
of high energy physicists. If you are suffering from this
problem and the Zoloft isn't working, pay attention and
I will attempt to dispel your weltschmerz by rational argument.

\subsubsection{Trotsky was right}

As it turns out, somewhat surprisingly, the guy who had
it right was Trotsky. \index{Trotsky, Leon}
This great philosopher and humanist
was the first to point out that high energy physics is exciting,
and will continue to remain exciting, precisely because it exists in a
state of {\it permanent revolution}. As experiments probe higher energies
and smaller distance scales, and as our theoretical frameworks
struggle to produce a coherent explanation of what we see,
our fundamental view of the physical world and how to describe it
changes dramatically. High energy physicists are continually
engaged in a process of creating new frontiers, moving to those frontiers,
civilizing the rough elements, then pushing out to the next frontier.

\begin{figure}
\begin{center}
\epsfig{file=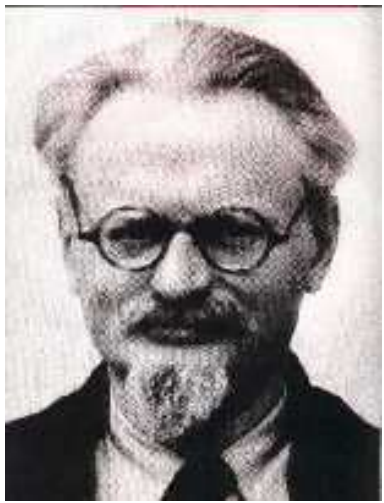,width=5cm}
\caption{Why is this man smiling?}
\label{fig:Trotsky}
\end{center}
\end{figure}

This rather obvious historical fact has lately been obscured by
two phenomena. One phenomenon is that particle physics
during the past decade was in
one of those recurring phases where theoretical assimilation of
earlier experimental findings dominates (to some extent) 
over new experimental surprises. This gives the
superficial impression that not much is happening, when in
fact during this period our understanding has increased dramatically,
such that our basic theoretical frameworks are remarkably different
from what they were 10 or 15 years ago.

The second phenomenon was the cancellation of the SSC,
and the concomitant popularization of the high energy physics version
of Horganism. \index{Horganism}
The idea here is that we have slipped into an era
of diminishing returns for our investments in basic science.
High energy physics, in this view, is becoming increasingly expensive and
complicated to pursue, and as a practical matter the field will
die out completely in a decade or two. 

This dire prediction may indeed be correct, but for reasons rooted in
politics and sociology, not physics. Precisely because high energy
physics is constantly redefining itself, we have no idea what
it will be like a century from now. Advances of the 21st century
could easily rival those of the 20th century.
There is no reason to imagine that we are near the end of this
process, barring the complete collapse of our civilization.
Or, as Trotsky put it, our expectation is

\begin{quote}
{\bf Revolution whose every successive stage is rooted in the
preceding one and which can end only in complete liquidation.}
\end{quote}

\subsection{The Standard Model as an effective field theory}

Let me now be much more explicit about the present status of
particle physics, and about the process of getting to the
next iteration of our understanding. The key realization
here is that the Standard Model is an effective field theory.
As advocated in
Ken Wilson's pioneering work~\cite{Wilson83}, \index{Wilson, Ken}
quantum field
theories in general are nonperturbatively defined
as effective field theories valid below some explicit
ultraviolet cutoff $\Lambda$. The parameters of the effective Hamiltonian
can be regarded as encoding the effects of integrating out
the ultraviolet degrees of freedom above the cutoff scale,
including the high momentum modes of the light degrees
of freedom. Effective Hamiltonians can contain an
arbitrarily large number of operators, but at energies small
compared to the cutoff the effects of higher dimension
operators are suppressed by powers of energy divided by
$\Lambda$.

The Standard Model, of course, is famously renormalizable.
However, as emphasized by
Steven Weinberg~\cite{Weinberg96}, \index{Weinberg, S.}
the power-counting renormalizability of the Standard Model has
no particular physical significance. The physically important
sense in which the Standard Model is renormalizable is that
the ultraviolet divergences of the model are controlled by
gauge symmetries, such that counterterms exist to cancel
all infinities. In this sense effective gauge field theories
with higher dimension operators are also renormalizable.
Renormalizability is in no sense an indication that the
Standard Model is ``fundamental''. Our generic expectation
is that the full Standard Model Hamiltonian will turn out to
contain a number of higher dimension operators, whose effects
are too suppressed to have shown up in present experiments.
Indeed it is this expectation which makes precision low energy
experiments interesting as probes of new physics.

In light of the above, the enterprise of searching for new physics
can be described in a completely model-independent way.
The task for experiments at future accelerators (as well as at existing
facilities) is to address the following set of physics questions:

\begin{itemize}
\item The Standard Model is an effective field theory for physics
below some high energy cutoff $\Lambda$. What is the value of
$\Lambda$?

\item What are the relevant degrees of freedom in {\it the new
effective theory} at energies above $\Lambda$.

\item What are the symmetries and organizing principles
of this new effective theory?

\item What symmetries and organizing principles of the Standard
Model turn out to be artifacts of the ``low energy'' approximation?

\item Do the symmetries and organizing principles of the new effective
theory explain the parameters and parameter hierarchies of the
Standard Model (e.g. all the notorious mysteries of flavor)?

\item Does the new effective theory give any hints (e.g. higher
dimension operators, spontaneously broken symmetries) of new
physics at even higher scales?

\end{itemize}

\subsection{What is the scale of new physics?}

Let me elaborate further on the first of these questions, which
has obvious importance for making smart choices about future
accelerators. How do we go about determining $\Lambda$ for the
Standard Model? There seem to be three basic complementary methods.

\begin{itemize}
\item One method is to use high energy machines to search for evidence of new
degrees of freedom characteristic of the new effective theory
above the cutoff. This could take the form of new particles,
resonances, or collective effects. It could also show up as
evidence of compositeness, form factors, indications of symmetry
restoration, or of symmetry breaking. We might even see signals
telling us about new spatial degrees of freedom. 

So far all such searches have turned up negative, even at
the Tevatron, our highest energy collider. This indicates either
that the new degrees of freedom are somewhat obscured (by
Standard Model backgrounds and instrumental effects) or that
present experiments are not yet probing above the cutoff scale
$\Lambda$.

\item Another method is to search for evidence of higher dimension
operators in the effective Hamiltonian of the Standard Model
itself. At high energies this approach is not entirely distinct
from the previous method, but it can also be utilized in a variety
of lower energy experiments. It is important to realize here
that the symmetries and approximate symmetries of the
known contributions to the Standard Model may not be
respected by the full effective action. This motivates a
variety of searches for flavor changing neutral currents,
CP violation, lepton number violation, proton decay, etc.

So far, with the exception of the strong case for
neutrino oscillations, experiments of this type have not
produced compelling results for new physics.
This is an indication either that certain operators are
simply forbidden in the full effective action, or that
$\Lambda$ is sufficiently large that these ``irrelevant''
operators decouple rather efficiently in current experiments. 

\item The third method is to consider the Higgs sector,
\index{Higgs}
which is {\it more sensitive}
to $\Lambda$ for several reasons. The first is that,
in the case where the Higgs mass is large,
the Higgs becomes strongly coupled
at some high energy scale. $\Lambda$ is
certainly no larger than this scale.
Similarly, in the case where the Higgs mass is small,
the Higgs effective potential
becomes unstable at some high energy scale, also
providing an upper bound on $\Lambda$.
These limits are summarized in Figure~\ref{fig:higgbound},
where we see that they have only logarithmic sensitivity
to the high scale. Also shown is the current 95\% confidence level
upper bound on the Standard Model Higgs mass from electroweak precision data,
which is approximately 250 GeV. Combined with the lower
bound on the Higgs mass from the direct searches at LEP,
the net result is that $\Lambda$ is rather weakly bounded.

\begin{figure}
\begin{center}
\epsfig{file=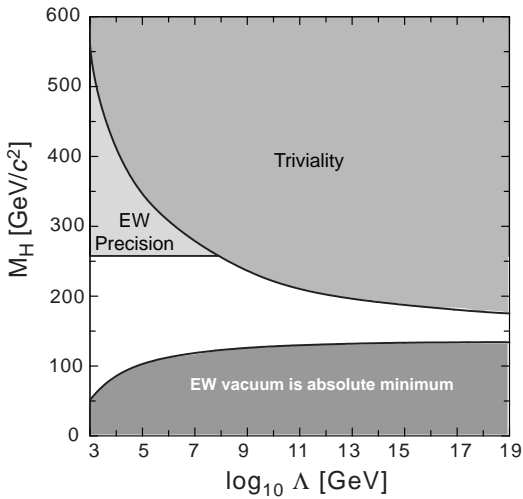,width=7cm}
\caption{The Higgs mass and upper bounds
on the cutoff scale. From C. Quigg, hep-ph/9905369.}
\label{fig:higgbound}
\end{center}
\end{figure}

A much stronger and more robust upper bound on $\Lambda$
is provided by consideration of {\it the Higgs naturalness
problem}. The Higgs has an unprotected scalar mass, whose
value is naturally of order the cutoff $\Lambda$. This
would lead us to the conclusion that $\Lambda$ not much more
than 250 GeV. A light Higgs can be arranged if, in the
new effective theory above $\Lambda$, the Higgs mass is
related to parameters which vanish in some symmetry limit.
However even in these cases it is hard to arrange a large
{\it hierarchy} between the electroweak scale (defined as
the Higgs vacuum expectation value $v = 246$ GeV)
and the cutoff $\Lambda$. 

\end{itemize}

The bottom line of this effective field theory analysis is that
there is a strong (and growing) tension in the Standard Model
between the Higgs naturalness/hierarchy problem, which wants
$\Lambda$ to be close to the electroweak scale, and the apparent
decoupling of new physics effects in current data.
The obvious resolution of this dialectic is that $\Lambda$
is only just out of reach of current experiments. An educated
guess would be:

$$
{\rm 500\ GeV} \lsim \Lambda \lsim {\rm 1\ TeV} 
$$

\medskip
I emphasize that although the numbers in this estimate
are a little soft, the physics input that goes into
it is very robust. Indeed it ultimately only depends
on our understanding of the Standard Model as an effective
quantum field theory.

\subsection{What could be out there?}

What will we find when we begin to probe the
new effective theory above the scale $\Lambda$?
Most theoretical speculation about the new
effective theory at high energies
involves {\it adding} things to the Standard Model:

\begin{itemize}
\item Add new particles: 4th generation, superpartners,
messenger sector, etc.

\item Add new symmetries: e.g. supersymmetry, etc.

\item Add new gauge interactions: e.g. technicolor, $Z^{\prime}$, etc.
\end{itemize}

\noindent However it is {\it just as likely} that at
higher energy scales we have instead (or in addition)
much more radical changes:

\begin{itemize}
\item Qualitatively new degrees of freedom: e.g. strings,
membranes, extra dimensions.
\item Symmetries of the Standard Model are broken: e.g. B and L violation.
\item ``Sacred principles'' of the Standard Model are violated!
This would not be the first time that sacred
cows got ground into hamburger.
\end{itemize}

\noindent Indeed, already in this century we have seen several examples
of sacred principles which turned out to be artifacts of
some approximation. It is worth reminding ourselves of these
history lessons:

\begin{enumerate}
\item Newtonian mechanics $\Leftrightarrow$ electromagnetism
$\rightarrow$ special relativity.

{\bf Lesson:} Galilean invariance is only an approximation, good at
low speeds.

\item Thermodynamics $\Leftrightarrow$ electromagnetism
$\rightarrow$ quantum mechanics.

{\bf Lesson:} Rayleigh's formula for blackbody emittance is only an
approximation, good at low frequencies.

\item Newtonian gravity $\Leftrightarrow$ special relativity
$\rightarrow$ general relativity.

{\bf Lesson:} Newtonian gravity is only an approximation, good for
weak gravitational fields and low speeds.
\end{enumerate}

{\bf A conservative view} is that all of
the following
theoretical assumptions/frameworks may break down
under certain conditions at certain energy scales:

\begin{itemize}
\item The assumption that the fundamental dynamical entities are
point-like particles.
\item Relativistic
quantum field theory (and the associated ideas of locality, microcausality,
CPT invariance).
\item General relativity.
\item Quantum mechanics. 
\end{itemize}

\noindent It is unfortunate that these possibilities are today largely ignored
by both theorists and experimenters. Of course we should not
ascribe every burp in the data to a breakdown of microcausality,
but neither should we assume that all of our current paradigms
will remain sacrosanct indefinitely. It is amusing to recall in
this regard that Werner Heisenberg, \index{Heisenberg, W.}
in 1939, seriously suggested that
quantum mechanics breaks down at an energy scale around 1 GeV.
Nowadays anyone who questions the universal validity of quantum
mechanics is (usually correctly) labelled as a crank.

Although string theory has not (yet) done a good
job of matching to the Standard Model at low energies, it has proven
to be a great exercise for both organizing and liberating our thinking
about new physics. For example:

\begin{itemize}
\item If string theory is correct, \index{string theory}
both general
relativity and quantum field theory break down at
some energy scale $M_s$. We don't know what this
string scale is; it's lower bound, set by experiment,
is about 1 TeV.

\item If string theory is correct, the fundamental
physical entities are not quarks and leptons, but
a whole collection of particle-like, string-like, and
membrane-like objects.

\item Furthermore these objects propagate in
a 9+1 or 10+1 dimensional spacetime.
\end{itemize}

\subsection{Model-independent conclusions}

Without making any particular assumptions about what new
physics is out there, we can now draw some
important conclusions with a high degree of confidence:

\begin{itemize}
\item There is {\it a whole new effective theory} waiting to
be explored at the TeV scale.

\item The new physics will be {\it rich, surprising, confusing,
and take a long time to untangle.}

These conclusions also imply the following:

\item To explore the new theory you will want high energies,
reasonable luminosities, and reasonable detectors.

\item To understand the new physics,
you will also want detailed studies, for which
you need excellent luminosities and excellent detectors.

\item You will need detailed studies not only to unravel
the new effective theory, but also to give you hints about
physics at even higher scales.
\end{itemize}
 
\section{Future accelerators}

In the second half of my talk I would like to discuss
the physics driving various proposals for future accelerators.
\index{accelerators, future}
To be concrete and focus our thinking I have summarized
these proposals below according to
my opinion of what we might have and when.
The dates of course are only estimates, but the groupings
are important. It is important in thinking about future machines
to make a clear distinction between those which are intended
to operate in the LHC era, and those which are clearly post-LHC
successors. It is also important to discriminate between those
proposals which extend the energy frontier, and those which
would operate within the energy frontier defined by the LHC.
I have also made note of possible upgrades of the Tevatron, LHC,
and linear $e^+e^-$ collider (LC),
since recycling is likely to become increasingly popular
in a difficult funding climate.

Given the very exciting prospect for future physics discoveries
advocated in the first half of this talk, it is rather straightforward
to make {\bf the physics case} for machines of the LHC era.
I emphasis that my summary of the physics needs does not factor in
the dollar cost or political cost of various machines, neither do
I address the question of whether we can attract sufficient human
resources to pursue several big projects simultaneously. I will
also touch briefly on machines of the post-LHC era. 

\clearpage
\noindent {\large\bf 2006 -- 2012: The LHC Era}
\begin{itemize}
\item {\bf LHC}: $\sqrt{s}$$=$$14$ TeV, $\cal{L}$$=$$10^{33}$ to $10^{34}$.
\item {\bf LC}: $\sqrt{s}$$=$$350$ GeV to 1 TeV,
$\cal{L}$$=$$10^{34}$ish.
\item {\bf $\nu$ factory}: 1 millimole of muons per year.
\item {\bf upgraded Tevatron?}: $\sqrt{s}$$= 4-6$ TeV.
\end{itemize}

\bigskip
\noindent {\large\bf 2013 -- 2025: Within the Energy Frontier}
\begin{itemize}
\item {\bf stretch LC}: $\sqrt{s}$$=$1.5 TeV.
\item {\bf $\gamma\gamma$, $e^-e^-$}: piggyback on LC.
\item {\bf First Muon Collider}: Higgs factory? Heavy Higgs factory?
\end{itemize}

\bigskip
\noindent {\large\bf 2013 -- 2025: Extending the Energy Frontier}
\begin{itemize}
\item {\bf upgraded LHC?}: $\sqrt{s}$$=$?.
\item {\bf CLIC}: $\sqrt{s}$$=$$3 - 5$ TeV, $\cal{L}$$=$$10^{35}$.
\item {\bf High Energy Muon Collider}: $\sqrt{s}$$=$$3-4$ TeV,
potential for 10 to 15 TeV.
\item {\bf VLHC}: $\sqrt{s}$$=$$100-200$ TeV, $\cal{L}$$=$$10^{35}$. 
\end{itemize}

\bigskip

\subsection{What is the physics driving the LHC?}

You are supposed to know this already! The LHC will advance the energy
frontier by roughly a factor of five over present experiments.
This should be amply sufficient to probe the new effective theory
above the cutoff $\Lambda$, discover most of its degrees of freedom,
symmetries, and organizing principles. Looking back, we should be 
able to get a fundamental understanding of the mechanism of electroweak
symmetry breaking, and perhaps shed light on flavor problems or other
mysteries of the Standard Model.

\subsection{What is the physics driving the LC?}

The various linear collider proposals involve
machines which will operate within the energy frontier
established by the LHC. I have attempted here to summarize
the important physics needs driving these proposals: 

\begin{itemize}
\item Higgs physics is golden.

\item The LHC won't be sufficient to unravel
the new physics at the TeV scale.

\item The LC has unique capabilities to divine
new physics at even higher scales.

\end{itemize}

\subsubsection{Higgs physics is golden}

Precision measurements are already telling us that either the
Higgs is light
(mass less than about 250 GeV ~\cite{LEPEWWG99}),
or new physics is misleading us!
Even if the Higgs turns out not to be Standard Model-like, it is still
likely to be discovered at either LEP, the Tevatron,
or the LHC. However, the
discovery of a $b\bar{b}$ invariant mass bump (say) in some data
sample will be only the beginning of understanding the Higgs.
As soon as a discovery is made, a main priority of high
energy physics will be to answer the following
questions~\cite{Carena99}: \index{Carena, M.}\index{Mrenna, S.}
\index{Wagner, C.}

\begin{enumerate}
\item Is this the Higgs of electroweak symmetry breaking?

\item Is this the Higgs associated with the generation
of fermion masses?

\item Is this the only Higgs?
\end{enumerate}

\begin{figure}
\begin{center}
\epsfig{file=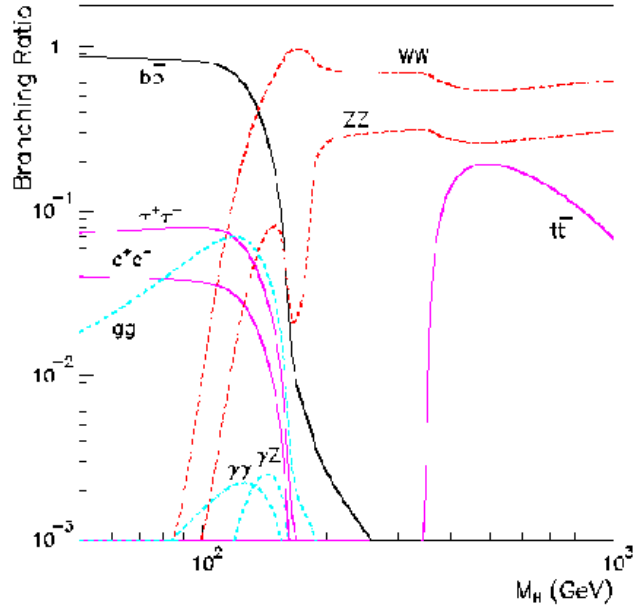,width=9cm}
\caption{Branching fractions versus mass for a
Standard Model Higgs.}
\label{fig:higgbrs}
\end{center}
\end{figure}

\noindent In short, we will need to find out everything we possibly
can about the Higgs sector.
Higgs physics will be golden, will occupy us for
many years, and will require a large number of challenging
measurements.

For example, we will
need precise measurements of all the Higgs
branching fractions. Disentangling the Higgs branching
to taus, charm, and glue is a tall order, and we will
need sensitivity to the rare decay modes as well.
The LHC can do part of this job, but we will need
an LC (with good luminosity and detectors) to do the rest.
For a light Higgs a 350 or 500 GeV linear collider has enough energy reach.
Eventually we would want to look at $t\bar{t}H$ production
at higher energies.

Higgs physics will be interesting for a long time.
Thus even in the post-LHC era we may be very interested
in low energy machines which can refine our knowledge
of the Higgs sector. This includes the
$\gamma\gamma$ option for a linear collider, 
and an s-channel Higgs factory as First Muon Collider.

\subsubsection{The LHC won't be sufficient to unravel
the new physics at the TeV scale.}

The LHC experiments will do a lot, including a large number
of precision measurements.
But, as I have argued, the new physics at the TeV scale will be
both rich and confusing. Mere prudence will demand that we
probe this new world with all the tools at our disposal.
A linear collider offers different sensitivities, polarization,
reduced backgrounds, better contained events, and even more precise
measurements. 

\noindent Examples:
\begin{itemize}
\item Untangling the neutralino and slepton sectors in
supersymmetry.
What variety of SUSY is it?

\item Deciphering virtual effects of extra dimensions.
Is your Drell-Yan anomaly due to spin 2 Kaluza-Klein
graviton exchange?
\end{itemize}

\subsubsection{LC precision measurements can pin down new physics scales}

A case study which illustrates this point was
recently made by
Ambrosanio and Blair~\cite{Ambrosanio99}. \index{Ambrosanio, S.}
\index{Blair, G}
They assumed that the
new physics is a minimal version of gauge-mediated supersymmetry,
and examined the question of whether experiments at a 500 GeV 
linear collider could measure the hidden sector supersymmetry
breaking scale $\sqrt{F}$.

This is a scenario in which the lightest neutralino is the
next-to-lightest-superpartner (NLSP),  and decays to
a Goldstino (which is not seen) plus a photon: $\tilde{\chi}^0_1$
$\rightarrow$$\gamma$G. The decay length of the NLSP,
$c\tau$, has only log sensitivity to the gauge mediation messenger scale,
but is proportional to the SUSY breaking scale $\sqrt{F}$:

$$c\tau_{\tilde{\chi}^0_1} \sim {F^2\over M_{\tilde{\chi}^0_1}^5}$$

\noindent This suggests a challenging physics program, in which we must
first determine that we have weak scale supersymmetry, that
we have gauge-mediated supersymmetry breaking, that this is
``minimal'' gauge mediation, and that the neutralino is the
NLSP. Having done all of this, we must then be prepared to
measure $c\tau$ of $\tilde{\chi}^0_1$ in the entire range from
10 microns to 30 meters, using various (overlapping) techniques:
\begin{itemize}
\item Projective tracking
\item 3D tracking
\item Photon pointing
\item Calorimeter timing
\item Statistical (counting single $\gamma$ versus 2 $\gamma$)
\end{itemize}

\noindent The conclusion of this study is that in such a scenario,
with an appropriate detector and
200 $fb^{-1}$ of integrated luminosity,
we could measure $\sqrt{F}$ to $\pm 5$\% at a 500 GeV linear collider.

\subsection{Why a Neutrino Factory?}

Neutrino oscillations are a strong hint that there is new
physics associated with scales in the $10^{10}$-$10^{16}$ GeV range,
or of brane-bulk physics in the case of
extra dimensions~\cite{Arkani-Hamed98}.
It will be a big job to pin down this new physics, and, in particular,
to link this new physics to anything else in the Standard Model
We will certainly need precise and overconstrained measurements of the
lepton mass matrix, just as we are now achieving for the CKM matrix
of the quark sector. Flavor problems are hard, and it seems highly
probable that we will need to build at least one new accelerator
facility optimized for neutrino physics. \index{neutrino factory}

\begin{figure}
\begin{center}
\epsfig{file=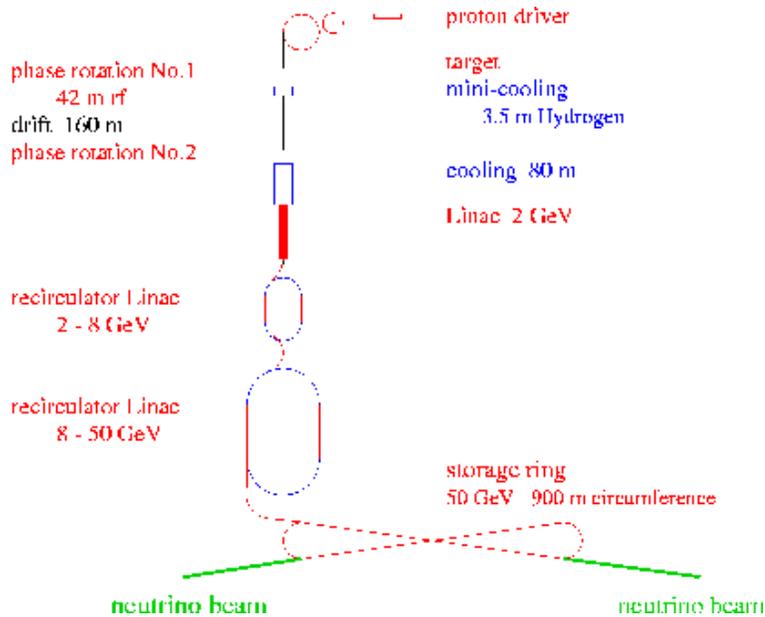,width=10cm}
\caption{Schematic for a muon storage ring
neutrino factory.}
\label{fig:nufact}
\bigskip
\end{center}
\end{figure}

Possible designs~\cite{Geer99} \index{Geer, S.}
of a muon storage ring neutrino factory
are currently under study~\cite{NorbertStudy,Autin99}.
the basic is to use a
muon storage ring as a source for very intense beams
of fairly high energy neutrinos.
The muon charge, momentum, and polarization determine
the neutrino composition and spectrum, thus the
initial characteristics of the neutrino beam will
be known with high confidence.
Many types of oscillation experiments can be
considered~\cite{Barger99}, \index{Barger, V.}
including very long baseline possibilities
such as Fermilab to Gran Sasso or CERN to Brookhaven.
This presents the opportunity for
{\it truely} international collaborations.
This is particularly so since there are
serious detector challenges to be overcome,
including building in the capability to discriminate
lepton flavors and measure their charges.

My belief that the neutrino factory can be brought
on line as an LHC era facility is based mostly on
its moderate size, and upon the growing interest and
enthusiasm exhibited in many quarters for this idea.
However a number of serious technical obstacles stand
in the way of this goal, and without an especially
aggressive R\&D effort my rosy scenario will not
come to pass. Since a neutrino factory appears
also to be the most practical path to a muon collider,
we have an especially strong motivation to pursue
this idea as vigorously as possible.

\subsection{Pushing the energy frontier}

The reality of modern high energy
physics is such that, if you want a new energy frontier
collider in 2020, you had better be doing serious R\&D
for it now. This puts us in something of a quandry, since
we don't yet know how to estimate
the next interesting energy scale.
From the arguments reviewed in this talk we
will certainly need a post-LHC energy frontier
machine, but beyond that general statement we
know almost nothing at this point in time.

In fact at the moment we cannot with confidence
answer even the most basic of questions. For
example, to explore the new physics of the
post-LHC era,
will a 3 or 4 TeV lepton collider, or an LHC
upgrade, be good enough, or do we need to push
straight to
a 10--15 TeV muon collider or 100--200 TeV VLHC?
Until we answer such questions we had better be
pursuing R\&D for all options.

The point here is that LHC+LC data is going to be essential for making
good decisions about post-LHC facilities. For example, 
LHC/LC data may indicate that, due to
the existence of extra dimensions, {\it the effective
Planck scale} is only about 3 TeV.
In this case hard scatterings at future
colliders may produce mostly black holes. 
This knowledge would certainly have
a major impact on your choices for $\sqrt{s}$, luminosity,
and detector design for post-LHC experiments!

\bigskip\medskip
\noindent{\large\bf A Final Thought}

\begin{quote}
It is much more likely that we will
fail to build new accelerators than that these
accelerators will fail to find interesting physics!
\end{quote}




\end{document}

%% file: LP99macros.tex
\def\beq{\begin{equation}}
\def\eeq#1{\label{#1}\end{equation}}
\def\eeqn{\end{equation}}

\def\beqa{\begin{eqnarray}}
\def\eeqa#1{\label{#1}\end{eqnarray}}
\def\eeqan{\end{eqnarray}}

\let\bar=\overbar

\def\Dslash{\not{\hbox{\kern-4pt $D$}}}
\def\dslash{\not{\hbox{\kern-2pt $\del$}}}

\def\msb{{\bar{\ssstyle M \kern -1pt S}}}




%% file: lykkenlp99.bbl
\begin{thebibliography}{99}

\bibitem{Jackson75}
J. D. Jackson, {\sl Classical Electrodynamics},
2nd Edition, John Wiley \& Sons, 1975.

\bibitem{TESLA98}
B. H. Wiik,
``The TESLA Project'',
Part. Accel. {\bf 62}, 43 (1998).

\bibitem{NLC96}
``Zeroth Order Design Report for the Next Linear Collider'',
SLAC-R-0474, May 1996.

\bibitem{VLHC97}
G. Anderson et al,
``Summary of the Very Large Hadron Collider Physics and Detector Workshop'',
hep-ph/9710254.

\bibitem{MUCOL99}
C. Ankenbrandt et al,
``Status of Muon Collider Research and Development and Future Plans'',
Fermilab-Pub-98-179,
Phys. Rev.ST Accel.Beams {\bf 2} 081001 (1999).

\bibitem{CLIC99}
J-P Delahaye et al,
``CLIC, a 0.5 TeV to 5 TeV $e^+e^-$ Compact Linear Collider'',
CERN/PS 99-005,
CERN/PS 99-062,
Acta Phys. Polon. {\bf B30}, 2029 (1999).

\bibitem{HWG99}
Draft report of the ``Physics at Run II'' Higgs Working Group,
available at http://fnth37.fnal.gov/higgs.html.

\bibitem{SHW99}
Draft reports of the ``Physics at Run II'' Supersymmetry/Higgs
Workshop, available at http://fnth37.fnal.gov/susy.html.

\bibitem{Dimopoulos98}
N. Arkani-Hamed, S. Dimopoulos, and G. Dvali,
Phys. Rev. {\bf D59} 086004 (1999).

\bibitem{Lykken99}
J. Lykken and L. Randall,
``The Shape of Gravity'',
hep-th/9908076.

\bibitem{Lykken96}
J. Lykken, Phys. Rev. {\bf D54} 3693 (1996).

\bibitem{Accomando99}
E. Accomando, I. Antoniadis, and K. Benakli,
``Looking for TeV Scale Strings and Extra Dimensions'',
hep-ph/9912287.

\bibitem{Cullen00}
S. Cullen, M. Perelstein, and M. Peskin,
``TeV Strings and Collider Probes of Large Extra Dimensions'',
hep-ph/0001166.

\bibitem{Wilson83}
K. G. Wilson, Rev. Mod. Phys. {\bf 55} 583 (1983).

\bibitem{Weinberg96}
S. Weinberg, {\sl The Quantum Theory of Fields},
Vol. II, Cambridge University Press, 1996.

\bibitem{LEPEWWG99}
The LEP Electroweak Working Group,
http://www.cern.ch/LEPEWWG/.

\bibitem{Carena99}
M. Carena, S. Mrenna, and C. E. M. Wagner,
``The Complementarity of LEP, the Tevatron, and the LHC
in the Search for a Light MSSM Higgs'',
hep-ph/9907422.

\bibitem{Ambrosanio99}
S. Ambrosanio and G. Blair,
``Measuring Gauge Mediated Supersymmetry Breaking
at a 500 GeV $e^+e^-$ Linear Collider'',
hep-ph/9905403.

\bibitem{Arkani-Hamed98}
N. Arkani-Hamed, S. Dimopoulos, and G. Dvali,
``Neutrino Masses from Large Extra Dimensions'',
hep-ph/9811448.

\bibitem{Geer99}
S. Geer, C. Johnstone, D. Neuffer,
``Design Concepts for a Muon Storage Ring Neutrino Source'',
Fermilab-Pub-99-121.

\bibitem{NorbertStudy}
FNAL Feasibility Study on a Neutrino Source Based on a Muon
Storage Ring,
http://www.fnal.gov/projects/muon\_collider/nu-factory/nu-factory.html.

\bibitem{Autin99}
B. Autin, A. Blondel, and J. Ellis (eds),
``Prospective Study of Muon Storage Rings at CERN'',
CERN-99-02, ECFA 99-197.

\bibitem{Barger99}
V. Barger, S. Geer, K. Whisnant,
``Long Baseline Neutrino Physics with a Muon Storage Ring Neutrino Source'',
hep-ph/9906487.

\end{thebibliography}
